\documentstyle[aps,prl,epsf]{revtex}

\begin{document}


\title{SIMULATION OF A NON-INVASIVE CHARGE DETECTOR FOR QUANTUM 
CELLULAR AUTOMATA}

\author{G. Iannaccone$^*$, C. Ungarelli, M. Macucci, E. Amirante, M.
Governale}

\address{
Dipartimento di Ingegneria dell'Informazione: 
Elettronica, Informatica e Telecomunicazioni \\
Universit\`a degli studi di Pisa, Via Diotisalvi 2, 
I-56126 Pisa, Italy}
\date{}
\maketitle

\begin{abstract} 

Information in a Quantum Cellular Automata architecture is encoded in
the polarizazion state of a cell, i.e., in the occupation numbers of
the quantum dots of which the cell is made up. Non-invasive charge
detectors of single electrons in a quantum dot are therefore needed,
and recent experiments have shown that a quantum constriction
electrostatically coupled to the quantum dot may be a viable solution.
We have performed a numerical simulation of a system made of a quantum
dot and a nearby quantum point contact defined, by means of depleting
metal gates, in a two-dimensional electron gas at a GaAs/AlGaAs
heterointerface. We have computed the occupancy of each dot and the
resistance of the quantum wire as a function of the voltage applied
to the plunger gate, and have derived design criteria for achieving
optimal sensitivity.

\end{abstract}

\section{Introduction}

Logic circuits based on the Quantum Cellular Automata (QCA) paradigm
offer an interesting alternative to traditional architectures used for
computation. \cite{lenttoug93,lenttoug93_2} The basic building block
of a QCA system is a cell made up of four or five quantum dots, and 
containing two electrons, which can align along one of the two
diagonals, giving rise to two possible polarization states. It has
been shown that cell polarization can propagate along a chain of cells
and that properly assembled two-dimensional arrays of cells allow the
implementation of logic functions. The results of any computation
performed with such arrays consist in the polarization state of some
output cells. This information must be read without perturbing it,
therefore a non-invasive charge probe is needed.

Recent experiments have shown that it is possible to detect a single
electron being added to a quantum dot by measuring the resistance of a
quantum point contact placed next to it. \cite{fielsmit93} The
electrostatic potential defining the constriction is modified by the
contribution of the additional electron, so that the transmission
coefficients and, consequently, the overall resistance are affected.

We have performed a numerical simulation of a system made of a
quantum dot and a nearby quantum point contact defined, by means of
depleting metal gates, in a two-dimensional electron gas at a
GaAs/AlGaAs heterointerface. The occupancy of each dot and the
conductance of the quantum wire are calculated as a function of the
voltage applied to the plunger gate for a few initial wire
resistances. The details of the simulations are described in
Section II, while the results are discussed in Section III.

We show that the highest sensitivity for charge detection
is obtained for bias voltages such that the initial resistance 
of the quantum wire is rather high, of the order of a hundred kiloohm,
when conduction through the quantum wire is substantially in the tunneling
regime, so that even extremely small variations of the confinement potential
(such as those due to the addition or removal of a single electron) produce
measurable variations of the resistance.

\section{Simulation}

The system we focus on is realized on the heterostructure sketched in
Fig. 1, consisting of an undoped GaAs substrate, an undoped 20~nm-thick
Al$_{0.36}$Ga$_{0.64}$As spacer layer, a Silicon delta doping layer
of $6 \times 10^{12}$~cm$^{-2}$, an undoped 10~nm-thick
Al$_{0.36}$Ga$_{0.64}$As layer, an undoped
5~nm-thick GaAs cap layer. 
An unintentional acceptor doping of
$10^{15}$~cm$^{-3}$ is considered in all undoped layers, while the delta
doping is modeled as a uniform doping on a 10~nm layer.

The gate configuration is shown in Fig. 2: the quantum dot is defined
by gates 1, 2, 3 and 4, and has a
geometrical area of 188 $\times$ 104 nm; the detector is the
constriction between gates 4 and 5.
The bias voltages $V_i$ of gates $i$ ($i=1 \dots 4$) are -0.12~V,
while we have considered a few different voltages $V_5$ for gate 5,
between -0.15 and 0.2 V, corresponding
to different initial resistances of the quantum costriction (from 7.6
k$\Omega$  to  520~M$\Omega$).

A detailed simulation of the system would require the self-consistent
solution of the Schr\"odinger and Poisson equations on a
three-dimensional grid, in order to obtain the conduction band edge
and the electron density profiles in the simulation domain. Then,
the resistance of the quantum point contact could be evaluated by 
means of the recursive Green's function formalism, \cite{macugali95} 
using the potential landscape obtained from the Poisson-Schr\"odinger 
solver, while the charge contained in the quantum dot could be simply
obtained by integrating the electron density in the dot region.
In order to assess the functionality of the detector, the
voltage of the plunger gate must be swept towards more 
negative values, as to progressively deplete the quantum
dot, and the induced quantum wire resistance must be calculated.
The simulation sketched above should be therefore repeated a
large number of times, and would be prohibitivetely time consuming.

For this reason, we have chosen a less rigorous approach, which is,
however, much simpler
from the computational point of view. First, the Poisson equation is
solved on a 3D grid (65 $\times$ 65 $\times$ 65 points) with a
semiclassical approximation: the electron and hole densities are
determined from the density of states in the bulk and the Fermi-Dirac
distribution, while acceptors are considered fully ionized. Dirichlet
boundary conditions are enforced for the potential at the gate surfaces,
while Neumann boundary conditions are enforced at the exposed GaAs
surface, with a normal component of the field equal to 88.2
V/$\mu$m, i.e., the value computed with all the gates at 0 V.
The Fermi level at equilibrium
is pinned 5.25~eV below the vacuum level, which is the value that
provides the best fit of the pinch-off voltage with experiments.
\cite{yongjin} On the lateral boundary regions of the simulation
domain, Neumann boundary conditions with zero electric field are enforced.

The contribution to the potential from the charge in the 
dot is calculated by solving again the Poisson equation, using
the charge in the dot as the only source term, and is subtracted
from the previously calculated potential, in order to obtain
the confining potential for the electrons.

At this point the 3D problem is decoupled into a 2D problem at the
heterointerface plane, and a 1D problem in the vertical direction,
under the assumption that confinement in the vertical direction
is much stronger than that on the horizontal plane. The one-dimensional 
Schr\"odinger equation is solved in the vertical direction, and the energy of
the first subband is obtained with respect to the conduction
band bottom in GaAs at the heterointerface: by adding this
value to the conduction band edge of the heterointerface plane, we
obtain the confining potential for electrons in the 2DEG, which
is reported in Fig. 3 for the case of $V_5 = -0.16$~V.

A 2D Schr\"odinger-Poisson self-consistent solver is used to
determine the dot occupancy and the charge distribution.
\cite{macuhess93}
The electron-electron interaction is modeled consistently with the Neumann
conditions enforced at the exposed surface, i.e., with negative image
charges.
Finally, the confining potential in the quantum wire is calculated by
adding to the previously obtained confining potential the contribution
of the electron density in the dot computed with the Schr\"odinger-Poisson 
solver. 

When the plunger gate voltage is modified, instead of solving again
the 3D Poisson equation, we use a semianalytical method
\cite{davilark95} to evaluate the correction to the confining
potential on the plane of the 2DEG, assuming that the other charges in
the structure remain unchanged. 

\section{Results and discussion}

In Fig. 4 the results of the simulation for $V_5 = -0.16$~V  are shown:
for a plunger gate voltage $V_2$ of -0.61 V the dot is completely depleted
and the detector resistance is 12.9~k$\Omega$. As $V_2$ is raised in
steps of 10~mV the confining potential on the heterointerface plane
is lowered, therefore the number of electrons in the dot $N$ (dashed line)
progressively increases.
The detector resistance decreases for increasing plunger gate voltage
as long as $N$ is constant; Instead, when one electron is added 
to the dot, Coulomb repulsion rises the confining potential of the quantum
costriction, causing an increase of a few percent of the detector
resistance.

We have also studied the dependence of the detector sensitivity upon
the initial resistance of the quantum constriction. In Fig. 5 the
detector resistance is plotted as a function of the plunger gate
voltage for four different voltages applied to gate 5, ranging
from -0.15~V to -0.18~V: of course, the
lower the voltage applied to gate 5, the higher the initial detector
resistance. As it can be seen, a high sensitivity can be obtained if
conduction in the quantum wire is essentially in the tunneling regime,
as in the cases of Figs. 5(d) and 5(c), corresponding to an initial
resistance much higher than that
associated to a single propagating mode in the quantum wire (i.e., 12.728
k$\Omega$).
This is simply due to the fact that the transmission probability in the
case of tunneling is extremely sensitive to a variation of the
confining potential profile. For values of $V_5$ lower than -0.18~V
the quantum wire is practically pinched off.

In order to have a detectable current through the detector in
quasi-equilibrium condition, an initial resistance close to
a hundred kiloohm, as in the case of Fig. 5(c), seems the better solution, 
since it also provides a relative change in the detector resistance
larger than 10 \% when one electron is added.

In conclusion, the electrostatic coupling between the dot and the 
constriction seems to be a viable detection principle for quantum
cellular automata systems. A more complete simulation, including
the four dots and the associated detectors, would be useful to 
really verify whether the ``stray'' capacitive couplings
between a detector and the other dots can undermine 
correct detection.




\section{Aknowledgments}

This work has been supported by the ESPRIT Project 23362 QUADRANT
(QUAntum Devices foR Advanced Nano-electronic Technology).
The authors wish to thank J. Martorell for making available
the code implementing the method of Ref. \cite{davilark95} .

\begin{figure}
\caption{Layer structure consisting of a GaAs substrate, 20 nm undoped AlGaAs spacer, Si delta
doping layer ($ 6 \times 10^{12}$~cm$^{-2}$), 10 nm undoped AlGaAs, 5
nm GaAs cap layer}
\end{figure}

\begin{figure}
\caption{Gate configuration defining the quantum dot and the detector
(dimensions are in nm).
Gate 2 is the plunger gate; the voltage applied to gate 
5 modulates the resistance of the detector.}
\end{figure}

\begin{figure}
\caption{Confining potential for the electrons in the plane of the GaAs-AlGaAs 
heterointerface.} 
\end{figure}

\begin{figure}
\caption{Detector resistance (solid line) and number of electrons in the
dot (dashed line) as a function of the voltage applied to the plunger 
gate}
\end{figure}

\begin{figure}
\caption{Detector resistances as a function of the voltage
applied to the plunger gate for four different values of the
voltage $V_5$. For $V_5$ lower the -0.18~V the quantum wire has
negligible conductance.}
\end{figure}

\newpage

\LARGE

\centerline{
\epsfxsize=14cm
\epsffile{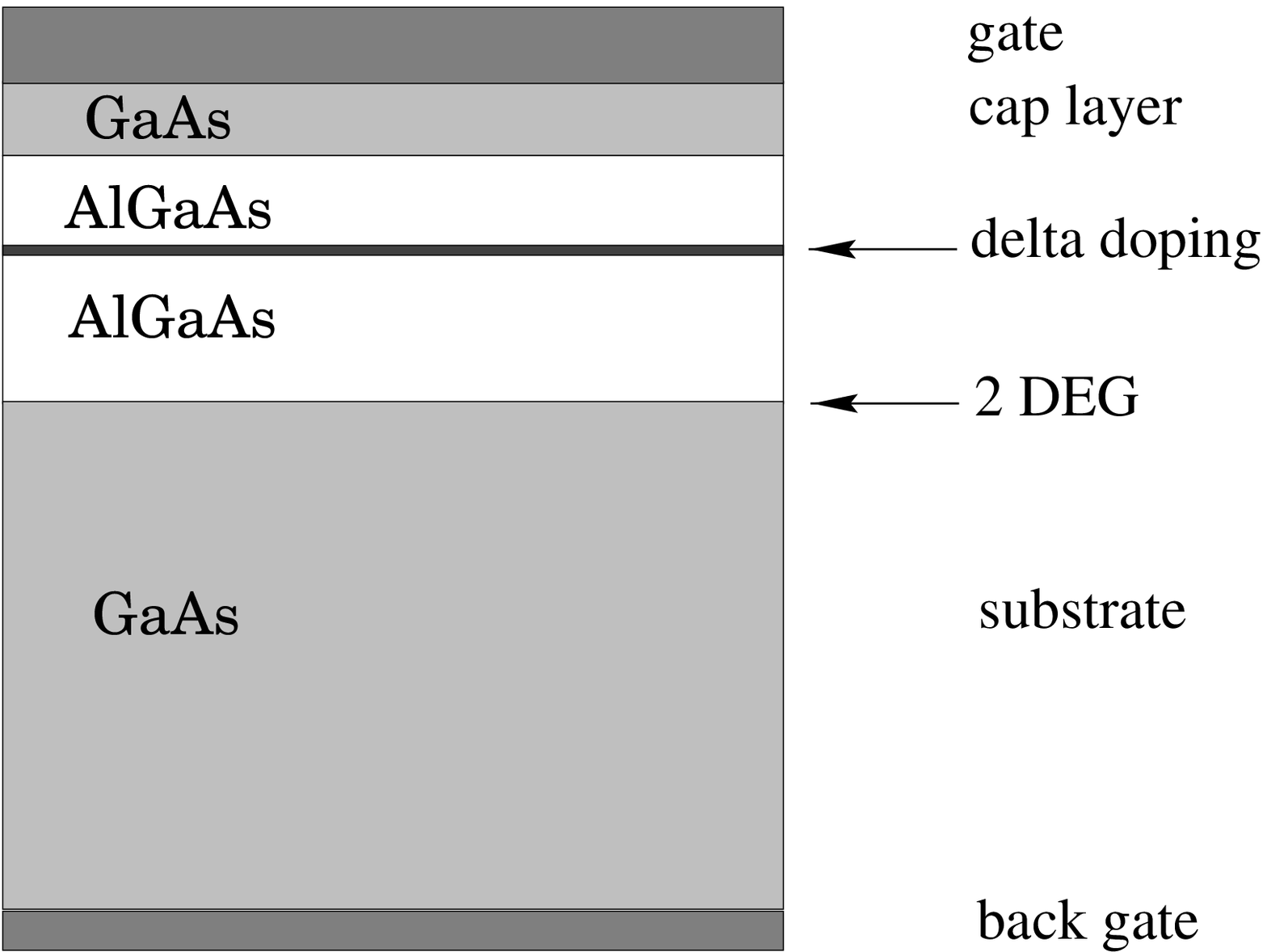}
}

\vspace{1cm}

Fig. 1

G. Iannaccone et al.

\newpage

\centerline{
\epsfxsize=11cm
\epsffile{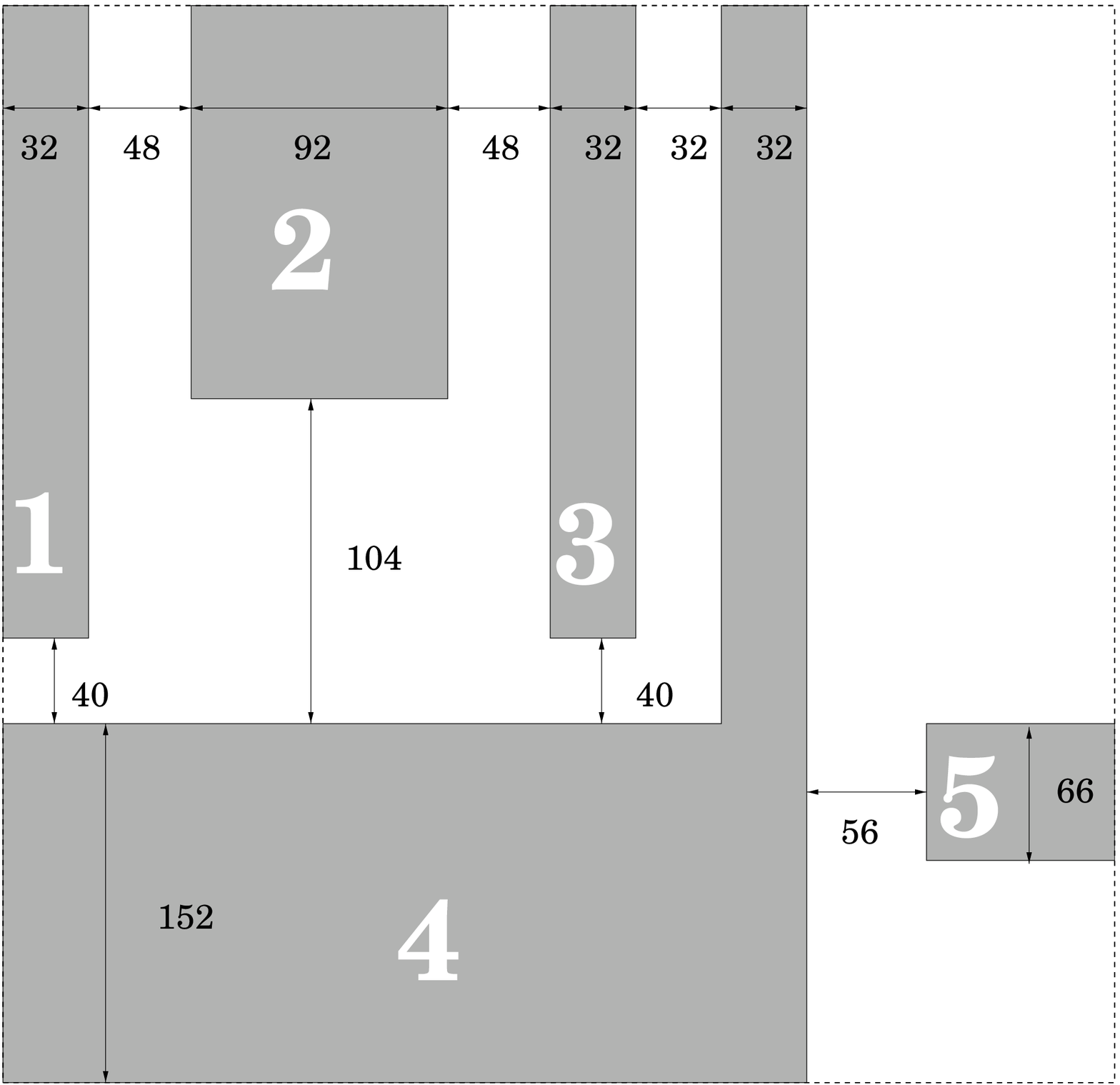}
}

\vspace{1cm}

Fig. 2

G. Iannaccone et al.

\newpage

\centerline{
\epsfxsize=15cm
\epsffile{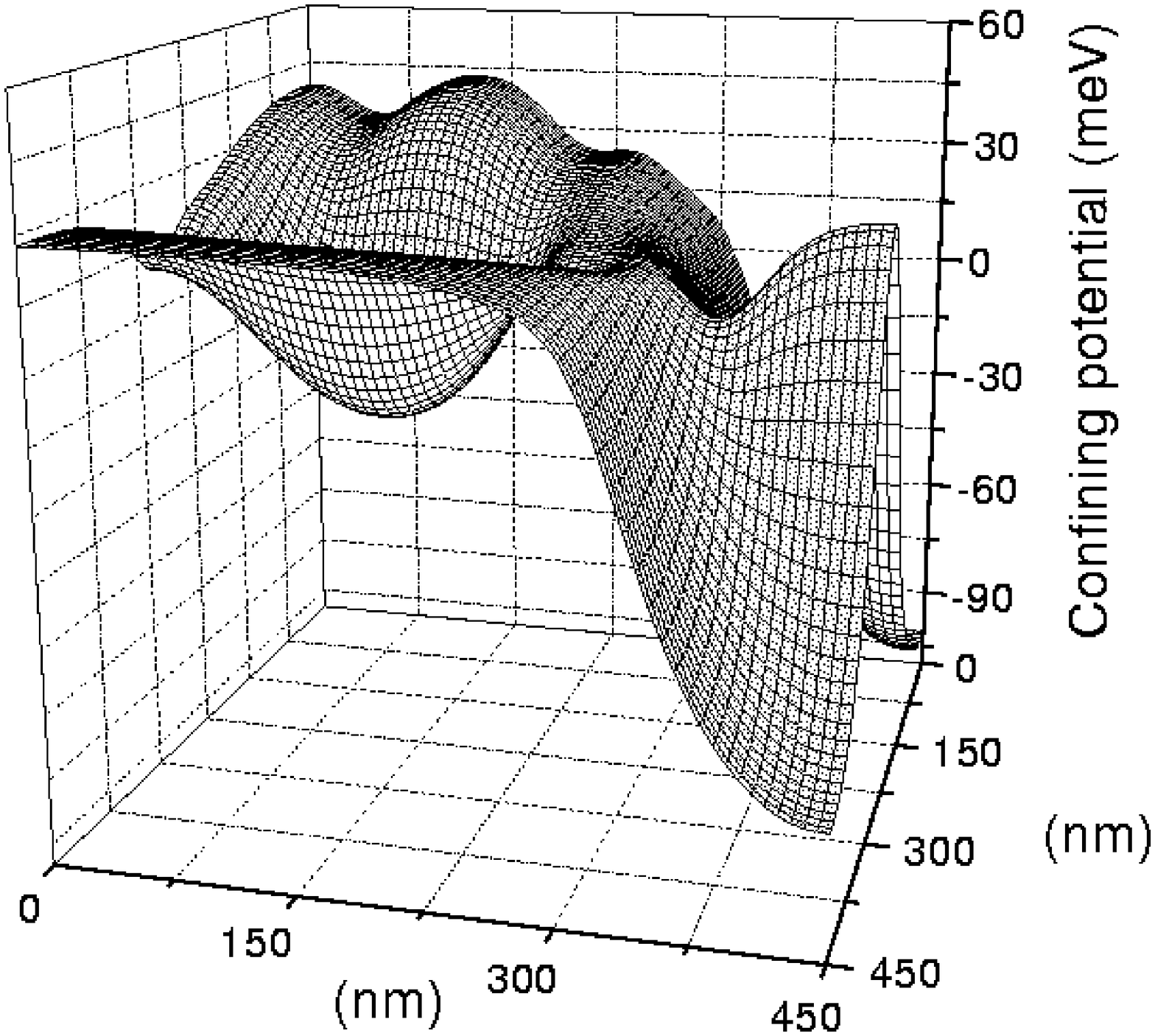}
}

\vspace{1cm}

Fig. 3

G. Iannaccone et al.

\newpage

\centerline{
\epsfxsize=15cm
\epsffile{figdet16}
}

\vspace{1cm}

Fig. 4

G. Iannaccone et al.

\newpage

\centerline{
\epsfxsize=12cm
\epsffile{figurona}
}

\vspace{1cm}

Fig. 5

G. Iannaccone et al.

\end{document}